\documentstyle[12pt]{article}
\def \k{\kappa}
\def \l{\lambda}
\def \pa{\partial}
\def \r{{\bf R}^2}
\def \A{A_0^{\k,q}}
\def \tA{\alpha ^{\k,q}}
\def \Li{\lim_{\k , q \to \infty }}
\def \Lint{\lim_{\k , q \to \infty }\int_{{\bf R}^2}} 

\def \su{\sum_{j=1} ^m}
\def \v{v^{\k,q}}

\def \intr{\int_{{\bf R}^2}}

\def \g{\gamma}
\def \e{\epsilon}
\def \E{{\cal F}}
\def \D{\Delta}
\def \vi{e^{v^i+f}}
\def \vii{e^{v^{i-1}+f}}

\def \bb{\begin{equation}}
\def \ee{\end{equation}}
\def \bbs{\begin{eqnarray}}
\def \ees{\end{eqnarray}}
\def \bqrn{\begin{eqnarray*}}
\def \eqrn{\end{eqnarray*}}
\newcommand{\qed}{\vrule width0pt\hfill \raisebox{-.3ex}{
   \frame{\phantom{\vrule height6pt width4pt depth4pt}}} \hspace*{-
7pt}}
\newtheorem{Thm}{Theorem}  
\newtheorem{Def}{Definition}
\newtheorem{Lemma}{Lemma}
\newtheorem{Corollary}{Corollary}
\newtheorem{Pro}{Proposition}
\newenvironment{pf}{\medskip\noindent{\it Proof:}\enspace}{\hfill \qed \newline \medskip}

\begin{document}
\title{Topological  Multivortices Solutions of the \\ Self-Dual 
Maxwell-Chern-Simons-Higgs System}
\author{
               Dongho Chae\\
        Department of Mathematics\\
        Seoul National University\\
           Seoul 151-742, Korea \\
          {\it e-mail: \ dhchae@math.snu.ac.kr} \\
                and \\
               Namkwon Kim\\
        Department of Mathematics\\
        POSTECH, Pohang 790-784, Korea\\
        {\it e-mail: \ nkkim@math.snu.ac.kr} }
\date{ }
\maketitle

\abstract{
We study existence and  various behaviors of topological multivortices 
solutions of 
the relativistic self-dual Maxwell-Chern-Simons-Higgs system. We first prove 
existence of general topological solutions by applying variational methods
to the newly discovered minimizing functional. 
Then, by an iteration method we prove existence of topological solutions
satisfying some extra conditions, which we call admissible solutions.
We establish asymptotic exponential decay estimates for these topological
solutions. We also investigate the limiting behaviors of the admissible
solutions as parameters in our system
goes to some limits. For the Abelian Higgs limit we obtain strong convergence
result, while for the Chern-Simons limit we only obtained that our
admissible solutions are weakly approximating one of the Chern-Simons solutions.  
}
\newpage
\section*{Introduction}

Since the pioneering works by Ginzburg and Landau on the superconductivity
there are
many studies on the Abelian Higgs system\cite{Taubes}, \cite{WangY}(and
references therein). In particular in \cite{Taubes} Jaffe and
Taubes established the unique existence of general finite energy 
multi-vortices solutions for
the Bogomol'nyi equations. (See also \cite{WangY} for more constructive existence proof
together with explicit  numerical solutions.)
More recently, motivated largely by the physics of 
high temperature superconductivity
the self-dual Chern-Simons system(hereafter Chern-Simons system) 
was modeled in \cite{H} and \cite{J}.(See \cite{D} for a general survey.)
The general existence theorem of topological solutions 
for the corresponding Bogomol'nyi equations
was established in \cite{Wang} by a variational method, and in \cite{Spruck}
 by an iteration
argument. 
For the nontopological boundary condition we have only  general existence result
for the radial solutions for vortices in a single point\cite{Spruck1}.  \\ 
We recall that in the Lagrangian of the Chern-Simons system there is 
no Maxwell term
appearing in the Abelian Higgs system, while the former includes
the Chern-Simons term which is not present in the later.
Naive inclusion of both of the two terms in the Lagrangian 
makes the system non self-dual(i.e. there is no Bogomol'nyi type equations
for the nontrivial global minimizer of the energy functional.)  
In \cite{Lee}, however, a self-dual system including both of the
 Maxwell and the Chern-Simons terms, so called 
(relativistic) {\it self-dual Maxwell-Chern-Simons-Higgs system} was successfully 
modeled using
the $N=2$ supersymmetry argument\cite{BLee}, \cite{Lee1}. 
It was found that we need 
extra neutral scalar
field to make the system self-dual. \\
In this paper we first prove general existence theorem for 
 topological 
multi-vortex solutions of the corresponding Bogomol'nyi 
equations of this system by a variational method. 
Then, using an iteration argument we constructively prove existence of
a class of solutions enjoying some extra conditions.
We call topological solutions satisfying these extra conditions the
{\it admissible topological solutions}.
We prove asymptotic exponential decay estimates for the
various terms in our Lagrangian for the general topological solutions. 
One of the most
interesting facts for the admissible topological solutions is 
that these solutions are really "interpolated"  between
the Abelian Higgs solution and the Chern-Simons solutions in the
the following sense: for fixed electric charge, when the Chern-Simons
coupling constant goes to zero, our solution converges  to the solution of
the Abelian Higgs system. The convergence in this case is very strong.
On the other hand, when both the Chern-Simons coupling constant and the
electric charge goes to infinity with some constraints between
the two constants, we proved that our solution is "weakly approximately"
satisfying the Bogomol'nyi equations for Chern-Simons system. 
In the existence proof for the admissible topological solutions, 
although we used iteration method in $\r$ directly, we could 
start iteration 
in a bounded domain with suitable boundary condition to obtain a solution
in that domain, and then enlarge this domain to the  whole of $\r$ as 
is done in
\cite{Spruck}, and \cite{WangY} in a much simpler case than ours. 
In this way it would be possible to obtain an explicit numerical solution. \\
The organization of this paper is following.\\
In the section 1 we introduce the action functional for the self-dual
Maxwell-Chern-Simons-Higgs system, and deduce a system of second order
elliptic partial differential equations which is a reduced version of the
Bogomol'nyi system.
In the section 2 by a variational method we prove a general existence of
topological solutions. Then, we introduce the notion of admissible
topological solution. In the section 3, 4 and 5 we prove existence of
admissible topological solutions by an iteration method.
In the section 5, in particular, we establish exponential decay estimates
for our solutions.
In the section 6 we prove strong convergence of the admissible topological
solutions to the Abelian Higgs solution.
The last section considers the Chern-Simons limit, and we prove that 
admissible topological solutions are weakly consistent to the Chern-Simons
equation in this limit. \\
({\it After finishing this work, we found that there was a study on the non
relativistic version of our model by Spruck-Yang in \cite{SY}.})

\section{Preliminaries}

The Lagrangian density for the (relativistic) 
self-dual Maxwell-Chern-Simons-Higgs system in 
$(2+1)$-D Lagrangian density modeled by  C. Lee {\it et al}\cite{Lee} is 
\bbs
{\cal L} 
     &=& -\frac{1}{4} F^{\mu \nu} F_{\mu \nu} 
         + \frac{\kappa}{4} \epsilon^{\mu \nu\lambda} F_{\mu \nu} A_\lambda 
                     + D_\mu \phi D^{\mu}\phi^* \nonumber  \\
     && + \frac{1}{2} \partial_\mu N \partial^{\mu} N - q^2 N^2 |\phi|^2 
        - \frac{1}{2} (q |\phi|^2 + \kappa N - q)^2    \label{Lag}
\ees
where $\phi$ is a complex scalar field, $N$ is a real scalar field, $A = (A_0, A_1, A_2)$ is a vector field,
$F_{\mu \nu} = \partial_\mu A_\nu - \partial_\nu A_\mu$, 
$D_{\mu} = \pa_\mu -i q A_{\mu}$, $\mu =0,1,2$,
$\pa_0 = \frac{\pa}{\pa t}$, $\pa_j =\frac{\pa}{\pa x_j}$, $j=1,2$, 
$q > 0$ is the charge of electron, 
 and $\kappa > 0$ is a coupling constant 
for Chern-Simons term. The action functional for this system is given by
\bb 
{\cal A} =\int _{{\bf R}^3} {\cal L} dx.\label{action} 
\ee 
The static energy functional for the above system is
\bbs
{\cal E}
&=& \int_{\r} \left\{ |D_0 \phi|^2 +|D_j \phi|^2 +\frac{1}{2}F_{j0} ^2 +
\frac{1}{2} F_{12} ^2 +\frac{1}{2} (\partial _j N)^2 \right.\nonumber \\
 && \ \ \ \ \left. +q^2 N^2 |\phi|^2 +\frac{1}{2}
       (q |\phi|^2 + \kappa N - q)^2 \right\}dx. \label{ener}
\ees
with the Gauss law constraint
\bb
(-\D +2 q^2 |\phi|^2 )A_0 = -\k F_{12}. \label{Gauss}
\ee
This  is the Euler-Lagrange equation with variation of the action taken
with respect to $A_0$.
Integrating by parts, using (\ref{Gauss}), we obtain from the energy functional 
\bbs
{\cal E}
&=& \int_{\r} \left\{ |(D_1 \pm i D_2)\phi |^2 +|D_0\phi \mp iq\phi N|^2
   +\frac{1}{2} (F_{j0} \pm \partial _j N )^2 \right. \nonumber \\
 && \ \ \ \ \left.+ \frac{1}{2}| F_{12} \pm (q|\phi|^2 +\kappa N -q)|^2 
\right\}dx
  \pm q\int_{\r} F_{12}dx. 
\ees
This implies the lower bound for the energy
\[ {\cal E} \geq q \left|\int_{\r} F_{12} dx\right|, \]
which is saturated by the solutions of the equations(the Bogomol'nyi equations)
 for $(\phi, A, N)$
\bbs
&& A_0 = \mp N \label{Bogo1}\\
&& (D_1 \pm i D_2) \phi = 0 \label{Bogo2}\\
&& F_{12} \pm (q|\phi|^2 + \kappa N - q) = 0 \label{Bogo3}\\
&& \Delta A_0 = \pm (\kappa q (1 - |\phi|^2) + 
\kappa^2 A_0 )+ 2q^2 |\phi|^2 A_0 \label{Bogo4} 
\ees
where the upper(lower) sign corresponds to positive(negative) values
of $\int_{\r} F_{12} dx$, and (\ref{Bogo4}) follows from the Gauss law
combined with (\ref{Bogo3}). \\
If  $(\phi, A, N)$ is a solution that makes ${\cal E}$ finite, then
either
\[ \phi \rightarrow 0 \ \ \ \mbox{and} \ \ \ \ N= -A_0 \rightarrow 
\frac{q}{\kappa}, \]
or
\[ |\phi|^2 \rightarrow 1 \ \ \ \mbox{and} \ \ \ \ N= -A_0 \rightarrow 0 \]
as $|x| \rightarrow \infty$.
The former is called non-topological, and the latter is called topological.
In this paper we are considering only topological boundary condition. 
We set  
\[
\phi = e^{\frac{1}{2} (u + i\theta)},\quad \theta = 
\sum_{j=1}^m 2n_j \arg (z - z_j), \quad
n_j \in Z^+ ,\]
where $z= (x,y)$ is the canonical coordinates in $\r$, and  each
$z_j=(x_j,y_j)$ is a zero of $\phi$ with winding number 
 $n_j$, which corresponds to
the multiplicity of the $j-$th vortex.
After similar reduction procedure similar to \cite{Taubes}, 
we obtain the equations(we have chosen the upper sign.)
\bbs
\Delta u &=& 2q^2 (e^u - 1) - 2q\kappa A_0 + 4\pi \su n_j\delta(z-z_j) \label{ueq} \\
\Delta A_0 &=& \kappa q (1 - e^u) + (\kappa^2 + 2q^2 e^u) A_0 \label{rawAeq}
\ees
with the boundary condition (\ref{bc}).
We define
\[
  f = \su n_j\,\, \ln\left(
              \frac{|z - z_j|^2}{1 + |z - z_j|^2}\right),   
\]
and we set $u = v + f$ to remove the singular inhomogeneous term 
in (\ref{ueq}).  Then (\ref{ueq}) and (\ref{rawAeq}) become
\bbs
\Delta v &=& 2q^2 (e^{v+f} - 1) - 2q\k A_0 + g, \label{veq}\\
\Delta A_0 &=& \k q (1 - e^{v+f}) + (\k^2 + 2q^2 e^{v+f}) A_0 \label{Aeq}
\ees
with the boundary condition
\begin{equation}
\lim_{|z|\rightarrow \infty} v= 0, \ \ \
\lim_{|z|\rightarrow \infty} A_0 =0, \label{bc}
\end{equation}
where
\[
g  = \su \frac{4n_j}{(1 + |z - z_j|^2)^2} 
\]

\section{Existence of a Variational Solution}

Solving (\ref{veq}) for $A_0$, and substituting this into (\ref{Aeq}), we 
obtain 
\bbs 
\Delta^2 v &-& (\k^2 + 4q^2e^{v+f})\Delta v 
        + 4q^4e^{v+f}(e^{v+f}-1) \nonumber \\
    &=& 2q^2|\nabla(v+f)|^2e^{v+f} -4q^2ge^{v+f} + \Delta g -\k^2 g
               \label{vari}
\ees
If we formally set $\k = 0$ in this equation, then we have
\[ 
\Delta ( \Delta v - 2q^2(e^{v+f} -1) - g) = 0 
\]
which, if we ask $v\in H^2 (\r )$, recovers
the Abelian Higgs system studied in \cite{Taubes}. 
On the other hand, if we take the limit $\k, q \rightarrow \infty$ with
$q^2/\k=l$ fixed number, then after formally dropping the lower order
terms in $o(q)$ and $o(\k)$, we obtain
\[ \D v =4l^2 e^{v+f} (e^{v+f} -1 )+g. \]
This is the equation corresponding to the pure Chern-Simons system studied
in \cite{Spruck}, \cite{Wang}, etc. \\
Later in this paper we provide rigorous justifications for these
two limiting behaviors of the solutions.
By a direct calculation we find that the equation (\ref{vari}) is a 
variational equation of the following functional.
\bbs \label{func}
{\cal F}(v) &=& \int \biggl[ \frac{1}{2}|\Delta v|^2 - (\Delta g -\k^2g )v 
             + 2q^4(e^{v+f}-1)^2  \nonumber \\
            && + \frac{1}{2}\k^2 |\nabla v|^2 
             + 2q^2e^{v+f}|\nabla (v+f)|^2 \biggr] dx
\ees
The above functional is well-defined in $H^2(\r)$ since $e^f |\nabla f|^2 
\in L^1(\r)$. We now prove existence of solution of (\ref{vari}) in $H^2(\r)$. 
Further regularity then $v \in H^2(\r)$ follows from the standard regularity
results for the nonhomogeneous biharmonic equations. 
\begin{Thm}
The functional (\ref{func}) is coercive, and weakly 
lower semi-continuous in $H^2(\r)$, and thus there is a global minimizer
of the functional (\ref{func}) in $H^2(\r)$.
\end{Thm}
\begin{pf}
If a sequence $\{v_k\}$ converges to $v$ weakly in $H^2(\r)$, 
$v_k \to v$ strongly in $L^{\infty}(B_R) $ and in  $H^1(B_R)$
for any ball $B_R=B(0,R) \subset \r$ by 
Rellich's compactness theorem. Thus we observe that 
to prove the lower semi-continuity of the functional (\ref{func}), 
it is sufficient
 to prove the lower semi-continuity of
$\int_{\r} e^{v+f}|\nabla (v+f)|^2 dx$.
We have
\bqrn
\liminf_{k\to \infty} \int_{\r} e^{v_k+f}|\nabla (v_k+f)|^2 dx 
&\geq& \liminf_{k\to \infty} \int_{B_R} e^{v_k+f}|\nabla (v_k+f)|^2 dx \\
        & =& \int_{B_R} e^{v+f}|\nabla (v+f)|^2 dx. 
\eqrn
Letting $R\rightarrow \infty$,  we obtain the desired weak 
lower semi-continuity. \\
On the other hand, we note that the  coercivity of $\E$ in $H^2(\r)$, 
is a simple corollary of the inequality:
\bb \label{coer}
\|v\|_{L^2(\r)}^2 \leq C ( 1+ \|\Delta v\|_{L^2(\r)}^2 
                      + \|e^{v+f}-1\|_{L^2(\r)}^2 ),
\ee
since we have 
\[ | \int (\Delta g - \k^2 g ) v dx | \leq \frac{C}{\e} 
         + \e \|v\|_{L^2(\r)}^2 \]
for any $\e > 0$, and by the Calderon-Zygmund inequality we have
\[\|D^2 v \|_{L^2 (\r )} \leq C \|\D v \|_{L^2 (\r )}. \]
For the proof of (\ref{coer}), we just recall the  inequality (4.10)
 in 
\cite{Wang}, which immediately implies;
\[ \|v\|_{L^2(\r)}^2 \leq C ( 1+ \|\nabla v\|_{L^2(\r)}^2 
                             + \|e^{v+f}-1\|_{L^2(\r)}^2 ).   \]
Now, for any $\eta>0$ we have
\[ \int |\nabla v|^2 dx = - \int v \Delta v dx 
              \leq \eta  \|v\|_{L^2(\r)}^2 
                  + C_{\eta} \| \Delta v \|_{L^2(\r)}^2.    \]
Taking $\eta$ small enough, (\ref{coer}) follows.   
This completes the proof of the theorem.
\end{pf}

\begin{Pro}
Let $(v, A_0)$ be any topological solution of (\ref{veq})-(\ref{Aeq}),
and $v_a ^q$ be the finite energy solution of the Abelian Higgs
system. Then the following
conditions are equivalent.
\begin{itemize}
\item[(i)] $A_0 \leq 0$
\item[(ii)] $v \leq -f$
\item[(iii)] $A_0\geq \frac{q}{\k} (e^{v+f} -1 )$
\item[(iv)] $v\leq v_a ^q $.
\end{itemize}
\end{Pro}
\begin{pf} \\
\noindent{\bf (i)$\Rightarrow$(ii),  (i)$\Rightarrow$(iv):} 
We assume  $A_0 \leq 0$.
Let $v_a^q $ be the solution of the Abelian Higgs system, i.e. $v_a^q$
satisfies
\bb
\Delta v_a^q = 2q^2 (e^{v_a^q + f} - 1) + g. \label{ah}
\ee
The existence and uniqueness of $ v_a^q \in H^2 (\r )\cap C^{\infty} (\r )$
satisfying $ v_a^q \leq -f$ is well-known\cite{Taubes}.
From (\ref{veq}) with $A_0 \leq 0$ we have
\[ \D v \geq 2q^2 (e^{v+f} -1 ) +g. \]
Thus,
\[ \D ( v_a^q -v) \leq 2q^2 (e^{ v_a^q f}-e^{v+f})= 2q^2 e^{\l +f} ( v_a^q -v) \]
by the mean value theorem, where $\l$ is between $v$ and $\v$.
By the maximum principle we have
\[v \leq v_a^q \leq -f. \] 
\noindent{\bf (ii)$\Rightarrow$(i):} We assume $v\leq -f$. From (\ref{Aeq}) we have
\[\D A \geq (\k ^2 +2 q^2 e^{v+f} )A .\]
Thus (i) follows from the maximum principle. \\
\noindent{\bf (i)$\Rightarrow$(iii):} We assume $A_0 \leq 0$.
Set $G= \frac{q}{\k} (1 - e^{v + f} )$. Then, we compute. 
\bqrn
\Delta G &=& -\frac{q}{\k} |\nabla (v + f)|^2 e^{v+f} - \frac{q}{\k}
              \Delta(v+f) e^{v + f} \\
 &\leq& -\frac{q}{\k} e^{v+f} [ 2q^2 (e^{v + f} - 1) 
                - 2q\k A_0 ] \\
 &=& 2q^2 e^{v+f} G +2q^2 A_0 
\eqrn
Thus, we have
\bqrn
\D (G+A_0 )&\leq& (\k^2 +2q^2 e^{v+f} )(G+A_0) +2q^2 A_0 \\
  &\leq& (\k^2 +2q^2e^{v+f} )(G+A_0 ).
\eqrn
Since $G\rightarrow 0$ as $|z| \rightarrow
\infty$, we have $G+A \geq 0$ by the maximum principle. \\
\noindent{\bf (iv)$\Rightarrow$(ii):} This is obvious, and included 
in the above proof. \\
\noindent{\bf (iii)$\Rightarrow$(i):} Assuming (iii), we have from (\ref{Aeq})
\[ \D A_0 \geq 2q^2 e^{v+f} A_0. \] 
Thus, (i) follows again by the maximum principle.
This completes the proof of the proposition.
\end{pf}
\ \\
\begin{Def}
We call a topological solution $(v, A_0)$ satisfying any one of the
four conditions in Proposition 1 by  an admissible topological solution.
\end{Def}

\section{Iteration Scheme}

In this section we construct an approximate 
 multi-vortices solution sequence of our Bogomol'nyi equations 
by an iteration scheme.
Later this approximate solution sequence  will be found to converge to an admissible
topological solution.
Our iteration scheme is similar  
to that of \cite{Spruck}, but is substantially extended in form.\\
\begin{Def} 
We set $v^0 = v_a^q$, $A_0^0 = 0$, where $v_a ^q$ is the finite energy
solution of the Abelian Higgs system. 
 Define
$(v^i, A_0^i) \in H^2(\r)\cap C^{\infty} (\r)$, 
$i \geq 1$ iteratively as follows: \\
First define $v^i$ from $(v^{i-1}, A_0^{i-1})$ by solving:
\bb        \label{vit}
(\Delta - d) v^i = 2q^2 (e^{v^{i-1} + f} - 1) - 2q\kappa A_0^{i-1} 
                + g - dv^{i-1},                          
\ee
and then define $A_0^i$ from $(v^i, A_0^{i-1})$ by solving:
\bb         \label{Ait}
(\Delta - \k^2 - 2q^2 e^{v^i + f} -d )A_0^i = \k q(1 - e^{v^i + f})
         - d A_0^{i-1}. 
\ee
Here, $d \geq 2q^2$ is a constant that will be fixed later.
\end{Def}

\begin{Lemma}
The scheme (\ref{vit})-(\ref{Ait}) is well-defined, and the 
iteration sequence $(v^i, A_0^i)$ satisfies the monotonicity, i.e.
\bqrn
 v^i  &\leq& v^{i-1} \leq \cdots \leq v^0 \leq -f \\
 A_0^i &\leq& A_0^{i-1} \leq \cdots \leq A_0^0 = 0.
\eqrn
\end{Lemma}
\begin{pf}
We proceed by an induction.
For $i=1$ we have from (\ref{vit}) 
\[ (\Delta - d) v^1 =  2q^2(e^{v^0+f}-1) + g -d v^0     \]
On the other hand $v^0$ satisfies
\[ \Delta v^0= 2q^2(e^{v^0+f}-1) + g. \]
Thus we have
\[  (\Delta - d)(v^1-v^0)=0. \]
From this we obtain $v^1=v^0 \in H^2(\r)\cap C^{\infty} (\r)$, and obviously 
\[ v^1 \leq v^0 \leq -f.\]
Now from (\ref{Ait}) for $i=1$ we have
\bb   \label{aaa}
(\Delta - \k^2 - 2q^2 e^{v^1 +f} -d )A_0^1 
              = \k q(1 - e^{v^1 + f}). 
\ee
Using the mean value theorem, we obtain
\[ \int_{\r} (1 - e^{v^1 + f})^2 dx\leq \int_{\r} (v^1 + f)^2 e^{\lambda +f} dx
\leq \int_{\r} (v^1 + f)^2 dx <\infty,\]
where $ v^1 <\lambda <-f $, and we used the fact $f\in L^2 (\r)$. 
We also have $0 \leq e^{v^1 +f} <1$.
Thus, by  the standard result of the linear elliptic theory 
the equation (\ref{aaa}) defines 
$A_0^1 \in H^2(\r)\cap C^{\infty} (\r)$. \\
Furthermore, since 
$ \k q(1 - e^{v^1 + f}) \geq 0,$ 
by the maximum principle applied to (\ref{aaa}) we have
\[ A_0 ^1 \leq A^0 _0 =0. \]
Thus  Lemma 1 is true for $i=1$.
Suppose the lemma is true up to $i-1$. Clearly
(\ref{vit})-(\ref{Ait}) define $(v^i, A_0 ^i )\in H^2 \cap C^{\infty} (\r )$
from $(v^{i-1}, A_0 ^{i-1} )$. We only need to observe that
\[
\int_{\r} (e^{v^{i} +f} -1)^2 dx  = \int_{\r} (v^{i} +f)^2 e^{\lambda +f}dx
<\infty \]
and 
\[ 0 \leq e^{v^{i} +f} \leq 1 \]
if $v^{i-1}\in L^2 (\r )$ and $v^{i-1} \leq -f$. 
We also have
\bqrn
(\Delta - d) (v^i - v^{i-1}) 
       &=& 2q^2(e^{v^{i-1} + f} - e^{v^{i-2} + f}) 
            - 2q \k (A^{i-1} - A^{i-2}) \\
       &&   - d(v^{i-1} - v^{i-2}) \\
       &\geq&  d (e^{v^{i-1} + f} - e^{v^{i - 2} + f} - (v^{i-1} - v^{i-2})) \\
       &=& d (e^{f+\lambda} - 1) (v^{i-1} - v^{i-2}),
\eqrn
where $v^{i-1} +f \leq \lambda \leq v^{i-2}+f$, and we used the mean value theorem.
Since 
$$ e^{w^i} \leq 1 \quad \mbox{and} \quad 
       v^{i-1} -v^{i-2} \le 0    $$
by induction hypothesis, we obtain 
\[ (\Delta - d) (v^i - v^{i-1}) \geq 0.  \]
Applying maximum principle again, we  have $v^i \leq v^{i-1}$.
On the other hand,
\bqrn
(\Delta - d) (A_0^i - A_0^{i-1}) &=& -\k q (e^{v^i + f} - e^{v^{i-1} + f}) \\
        && + (\k^2 + 2q^2 e^{v^i + f}) (A_0^i - A_0^{i-1})\\
        && + 2q^2 (e^{v^i + f} - e^{v^{i-1} + f}) A_0^{i-1} 
                       - d (A_0^{i-1} -A_0^{i-2})\\
        &\geq& (\k^2 + 2q^2 e^{v^i + f} ) (A_0^i - A_0^{i-1}),
\eqrn
where we used the assumption  that our lemma holds up to $i-1$, and 
$v^i \leq v^{i-1}$.
Therefore, by the maximum principle,
$$ A_0^i \leq A_0^{i-1} $$
Lemma 1 is thus proved.
\end{pf}

\begin{Lemma} 
The iteration sequence $(v^i , A_0^i)$ satisfies the inequality
\[ 
A_0^i \ge \frac{q}{\k} (e^{v^i+f} - 1).    
\]
for each $i=0,1,2, \cdots$.
\end{Lemma}

\begin{pf} 
We use induction again.
Lemma 2 is true for $i=1$.
We set $G^i = \frac{q}{\k} (1 - e^{v^i + f} )$.
Suppose Lemma 2 holds for $i-1$, then
\bqrn
\Delta G^i &=& -\frac{q}{\k} \nabla \cdot (\nabla(v^i + f) e^{v^i+f}) \\
         &=& -\frac{q}{\k} |\nabla (v^i + f)|^2 e^{v^i+f} - \frac{q}{\k}
              \Delta(v^i+f) e^{v^i + f} \\
         &\leq& -\frac{q}{\k} e^{v^i+f} (2q^2 (e^{v^{i-1} + f} - 1) 
                - 2q\k A_0^{i-1} + d(v^i - v^{i-1}))\\
         &\leq& 2q^2 e^{v^{i-1} + f}G^i
                + \frac{q}{\k} ( e^{v^i+f} - e^{v^{i-1} + f}
                - \frac{q}{\k}d e^{v^i + f} (v^i - v^{i-1})    \\
         &\leq& 2q^2 e^{v^{i-1} + f}G^i
                - \frac{q}{\k}d e^{v^i + f} (v^i - v^{i-1}) 
\eqrn
by $A_0^i \leq 0$, $i \geq 0$.
Therefore 
\bqrn 
(\Delta - d)(G^i + A_0^i) 
       &\leq& (\k^2 + 2q^2 e^{v^i + f} ) (A_0^i + G^i)\\
       && - d G^i -d A_0^{i-1}
          - \frac{q}{\k}d e^{v^i + f} (v^i - v^{i-1})  \\
       &\leq& (\k^2 + 2q^2 e^{v^i + f} ) (A_0^i + G^i) \\
       && - d ( A_0^{i-1} + G^i + \frac{q}{\k} (e^{v^i+f} - e^{v^{i-1} + f} ),
\eqrn
where we used the mean value theorem in the last step,and used the fact
$v^i \leq v^{i-1}$.
Rewriting it, we have
\bqrn
(\Delta - \k^2 - 2q^2 e^{v^i + f} -d ) (G^i + A_0^i)
        \leq -d ( A_0^{i-1} + G^{i-1}) \leq 0
\eqrn
by the induction hypothesis.  By maximum principle we have  \\
$A_0^i + G^i \geq 0$.
This completes the proof of Lemma 2.
\end{pf}

\section{Monotonicity of ${\cal F}(v^i)$}

In this section we will prove the following:
\begin{Lemma}
Let $\{v^i \}$ given as in Definition 2 and $\E (v)$ 
is given in (\ref{func}). We have 
\bb  \label{E}
\E (v^i) \leq \E(v^{i-1}) \leq \cdots \leq \E (v^0 ).
\ee
\end{Lemma}
To prove this we firstly begin with:
\begin{Lemma}  \label{G-b}
Let $(v^i, A_0^i)$ be as in Definition 2, then
\[ 
\int_{\r} \left[ |1-e^{v^i+f} + \frac{\k}{q} A_0^i| 
            + \frac{d}{2q^2} |v^i-v^{i+1}| \right] dx 
            = \frac{2\pi}{q^2} \su n_j
\]
for all $i \geq 0$.
\end{Lemma}
From Lemma 1 and 2 we have 
\[ 1-e^{v^i + f} + \frac{\k}{q} A_0^i, \;\; v^i - v^{i+1} \geq 0. \]
We only need to prove
\[ \int_{\r} \left[1-e^{v^i + f} + \frac{\k}{q} A_0^i 
              + \frac{d}{2q^2}(v^i - v^{i+1})\;\right] dx = 
                     \frac{1}{2q^2} \int_{\r} g \; dx.   \]
Fix $R > 0$. Integrating (\ref{vit}) over $B_R=\{|z|< R\}$, we obtain
\bbs       
&&\int_{B_R} \left[ (1-e^{v^i + f} + \frac{\k}{q} A_0^i) 
          + \frac{d}{2q^2}(v^i - v^{i+1}) \;\right] dx \nonumber \\
       && \quad = \frac{1}{2q^2} \int_{B_R} 
                 (g - \Delta v^{i+1}) \; dx 
                \label{G}
\ees
By divergence theorem
$$\int_{B_R} \Delta v^i = \int_{\partial B_R} 
                \frac{\partial v^i}{\partial r} d\sigma.$$
We note that $v^i \in H^1 (\r)$.
Thus
\bb \label{vbdry}
\int_{\partial B_R} |\nabla v^i| d\sigma \leq 
            \left(2\pi R \int_{\partial B_R} |\nabla v^i|^2 
                        d\sigma\right)^{1/2}
\ee
by H\"{o}lder's inequality.
Let 
\[  H(r) = \int_{\partial B_r} |\nabla v^i|^2 d\sigma,   \]
then
\[  
\int_{\r} |\nabla v^i|^2 \; dx  = \int_0^\infty H(r) dr < +\infty   
\]
Therefore there exists an increasing sequence of radii, 
$\{r_k\}_{k=1}^\infty$, such that
\[ 
\lim_{k\to \infty} r_k = +\infty,\mbox{and} \quad H(r_k) < \frac{o(r_k)}{r_k} 
\]
Otherwise, there exists $\epsilon > 0$ and $\tilde{r} > 0$ such that 
$H(r) > \frac{\epsilon}{r}$ for $r > \tilde{r}$, but then 
$\int_{\r} |\nabla v^i|^2 \; dx = \infty$.
Thus (\ref{vbdry}) implies
\[ 
\int_{\partial B_{r_k}} |\nabla v^i| d\sigma  
        \leq (2\pi r_k H(r_k))^{1/2} \leq (2\pi o(r_k))^{1/2}.   
\]
Therefore
\[
\lim_{k\to \infty} \int_{\partial B_{r_k}} |\nabla v^i| d\sigma = 0. 
\]
Choose $R = r_k$, and let $k\to \infty$ in (\ref{G}), then we have
\[ 
\lim_{k\to\infty} \int_{B_{r_k}} \left[1-e^{v^i + f} + \frac{\k}{q} 
           A_0^i + \frac{d}{2q^2} (v^i -v^{i+1}) \right] \; dx 
                                 = \frac{1}{2q^2} \int_{\r} g \; dx.
\]
This, together with
\[ \int _{\r} gdx =\su 8\pi n_j  \int_0 ^{\infty}
\frac{rdr}{(1+r^2)^2}=
 4\pi \su n_j \]
completes the proof of the lemma.
\qed

As a corollary of Lemma \ref{G-b}, we can get the following uniform bound.
\begin{Corollary} \label{S-b}
Let $(v^i,A_0^i)$ be as in  Definition 1, and define 
\bb \label{S}
S = \sup_{i\geq 1} \left(\intr \vi |\nabla (v^i+f) |^2 dx \right)^{\frac{1}{2}} 
\ee
Then $S \leq (4\pi \su n_j )^{\frac{1}{2}}$.
\end{Corollary}
\begin{pf}
Multiplying (\ref{vit}) by $e^{v^i+f}$ and integrating 
by parts, we have
\bqrn 
\intr \vi |\nabla(v^i + f)|^2 dx
         &=& \intr \biggl[ d(v^{i-1} - v^i) \vi  \\
         &&  + 2q^2 \vi(1 - \vii + \frac{\k}{q}A_0^{i-1})  \biggr] dx  \\
         &\leq& \intr \biggl[ d(v^{i-1} - v^i) 
	     + 2q^2 (1 - \vii + \frac{\k}{q}A_0^{i-1})  \biggr] dx   \\
         &\leq& 4\pi \su n_j
\eqrn
by Lemma \ref{G-b}.
\end{pf}
We now prove our main lemma in this section. \\
\ \\
\noindent{\it Proof of Lemma 3:}
From  (\ref{func}) we have
\bqrn
\E(v^{i-1}) - \E(v^i) &=& 
     \intr \biggl[ \frac{1}{2} |\Delta (v^i -v^{i-1})|^2 
          - \Delta v^i \Delta (v^i - v^{i-1})   \\
      &&  + (\Delta g -\k^2 g)(v^i - v^{i-1})
          + \frac{\k^2}{2}|\nabla(v^i - v^{i-1})|^2 \\
      &&  - \k^2 \nabla(v^i - v^{i-1}) \cdot \nabla v^i
          + II + III \biggr] dx,
\eqrn
where 
\bqrn
II  &=& - 2q^4 \biggl[ (e^{v^i +f }-1)^2 - (e^{v^{i-1} +f} - 1)^2 \biggr]   \\ 
III &=& -2q^2 \biggl[ e^{v^i +f}|\nabla (v^i+f)|^2 
            - e^{v^{i-1}+f}|\nabla(v^{i-1}+f)|^2  \biggr] .      
\eqrn
We also set
\[ I =  - \Delta v^i \Delta (v^i - v^{i-1}) + (\Delta g -\k^2 g)(v^i - v^{i-1})
          - \k^2 \nabla(v^i - v^{i-1}) \cdot \nabla v^i.              \]
Then 
\bbs    
\E(v^{i-1}) - \E(v^i) &=& \intr \biggl[ \frac{1}{2} |\Delta (v^i -v^{i-1})|^2
          + \frac{\k^2}{2}|\nabla(v^i - v^{i-1})|^2 \nonumber \\
          && + I + II + III \biggr] dx .  
             \label{Ediff}
\ees
We firstly estimate $I$.
From (\ref{vit}) we have
\[ A_0^{i-1} = \frac{1}{2q\k}\left(2q^2 (e^{v^{i-1}+f}-1)+g -d(v^{i-1} -v^i\right)
                - \Delta v^i ).                                       \]
Putting this into (\ref{Ait}) after substituting $i$ 
with $i-1$ in (\ref{Ait}), we have
\bbs
\Delta^2 v^i - (d +\k^2 + 2q^2 \vii )\Delta v^i
        + (d\k^2 + 2q^2 d \vii )v^i  \nonumber \\
      = (2q^2\vii -d)\Delta v^{i-1} + 2q^2 |\nabla(v^{i-1} +f)|^2 \vii 
                     \nonumber \\
        -4q^2g\vii + \D g - \k^2 g   + d(\k^2 + 2q^2 \vii)v^{i-1} 
                     \nonumber \\
        - 4q^4 \vii (\vii -1) - 2q\k d (A_0^{i-1} - A_0^{i-2})  
                     \label{vtot}.
\ees
Multiplying (\ref{vtot}) by  $v^i -v^{i-1}$, integrating 
by parts, we have
\bqrn
\intr \left[ \Delta v^i \Delta (v^i - v^{i-1}) 
     + \k^2 \nabla(v^i - v^{i-1}) \cdot \nabla v^i 
     - (\Delta g -\k^2 g)(v^i - v^{i-1}) \right] dx   \\
     = - \intr \biggl[ d|\nabla(v^i - v^{i-1})|^2 
       + (d\k^2 + 2q^2 d \vii)(v^i - v^{i-1})^2  \\
       + 2q\k d (A_0^{i-1} - A_0^{i-2})(v^i -v^{i-1})
       - IV - V \biggr] dx,
\eqrn
where we set 
\bqrn
IV &=& - 4q^4 \vii (\vii -1)(v^i -v^{i-1})     \\
V  &=&  2q^2 \biggl[ \vii (v^i -v^{i-1}) \D (v^i + v^{i-1}) \\
   &&   + |\nabla (v^{i-1} +f)|^2 \vii (v^i -v^{i-1}) 
        + 2g \vii (v^i - v^{i-1}) \biggr] .  
\eqrn
Recalling the definition of $I$ and observing 
$(A_0^{i-1} - A_0^{i-2})(v^i -v^{i-1}) \geq 0$, we get
\bb   \label{I}
I + IV + V \geq  d|\nabla(v^i - v^{i-1})|^2 
                   + (d\k^2 + 2q^2 d \vii)(v^i - v^{i-1})^2.
\ee
To calculate $I + II + III$, we observe 
\bqrn
II - IV = 4q^4(v^i-v^{i-1}) \biggl[ \vii(\vii - 1) 
                         - e^{\l +f}(e^{\l +f} -1)  \biggr] \\
        = 4q^4(v^i-v^{i-1})(v^{i-1} - \l) e^{\eta + f}(2e^{\eta +f} -1), 
\eqrn
where we used mean value theorem repeatedly with 
$v^i \leq \eta \leq \l \leq v^{i-1}$.
Thus
\bb              \label{II}
    II - IV \geq -4q^4 (v^i-v^{i-1})^2 \vii .      
\ee
Now we have
\bqrn
III &=& - 2q^2 [ (\vi - \vii )|\nabla (v^i+f)|^2  \\
    &&  + \vii ( |\nabla (v^i +f)|^2 - |\nabla (v^{i-1}+f)|^2 ) ] \\
    &=& - 2q^2 [ e^{\l +f}(v^i -v^{i-1})|\nabla (v^i+f)|^2   \\
    &&  + \vii \nabla(v^i -v^{i-1}) \cdot \nabla (v^i + v^{i-1} +2f) ] \\
    &=& VI + VII,                                           \\
V &=& 2q^2\vii (v^i - v^{i-1})[ \D (v^i + v^{i-1} + 2f)  \\
    && + |\nabla(v^{i-1} + f)|^2 ]                     \\
  &=&  VIII + IX,
\eqrn
where $v^{i-1} \geq \l \geq v^i$ by mean value theorem.
We now calculate $III - V = (VII - VIII) - IX + VI$. 
By integration by parts we obtain 
\bqrn
&& \int_{\r} [ \ VII - VIII \ ] dx \\ 
     &&\quad = 2q^2 \int_{\r}\left[ \vii (v^i -v^{i-1}) \nabla(v^{i-1}+f)\cdot
                  \nabla (v^i + v^{i-1} +2f)\right] dx \\
     &&\quad =  \int_{\r} [ X ]dx,    \\
&&  X - IX   = 2q^2 \vii (v^i - v^{i-1}) \nabla (v^i + f)
                 \cdot \nabla ( v^{i-1} + f)  \\
     &&\quad = XI.
\eqrn
Since $VI \leq 0$, we have
\bqrn
VI + XI &\geq& - 2q^2 (v^i - v^{i-1}) \biggl[ \vi \nabla (v^i+f) 
                 \cdot \nabla (v^i - v^{i-1})  \\
        &&     + (\vi - \vii ) \nabla (v^i+f)\cdot \nabla (v^{i-1}+f) \biggr] \\
        &\geq& - 2q^2 |v^i - v^{i-1}| \biggl[ \vi |\nabla (v^i+f)|
                    |\nabla (v^i - v^{i-1})|    \\
          &&   + |\vi - \vii||\nabla (v^{i-1}+f)|^2 \\
          &&   + |\vi - \vii| |\nabla (v^i - v^{i-1})| |\nabla (v^{i-1}+f)| 
                    \biggr]\\
        &\geq& - 2q^2 |v^i - v^{i-1}| \biggl[ (\vi |\nabla (v^i+f)| \\
          &&   +\vii |\nabla (v^i - v^{i-1})| |\nabla (v^{i-1}+f)|) \\
          &&   + e^{\l +f}|v^i - v^{i-1}| |\nabla (v^{i-1}+f)|^2 \biggr]
\eqrn
by the mean value theorem where we used the fact $|\vi - \vii| \leq \vii$ 
in the last step.
We use H\"{o}lder's inequality and interpolation inequality to obtain
\bqrn
\int_{\r} [ \ VI + XI \ ]dx 
       &\geq& - 2q^2 \left( \| \vi \nabla (v^i+f) \|_{L^2(\r)}
               \right. \\
&& +\|\left. \vii \nabla (v^{i-1}+f) \|_{L^2(\r)} \right) \\
          && \times  \| (v^i - v^{i-1})\|_{L^{\infty}(\r)}
                \|\nabla(v^i - v^{i-1})\|_{L^2(\r)}    \\
          &&  - 2q^2 \|v^i - v^{i-1}\|_{L^{\infty}(\r)}^2 
                       \intr \vi |\nabla (v^i+f) |^2 dx    \\
       &\geq& -4q^2C S \|v^i - v^{i-1}\|_{L^2(\r)}^{\frac{1}{2}} 
                       \|\D(v^i - v^{i-1})\|_{L^2(\r)}^{\frac{1}{2}} \\
          &&    \times \|\nabla(v^i - v^{i-1})\|_{L^2(\r)}    \\
          &&  -2q^2 C S^2 \|v^i - v^{i-1}\|_{L^2(\r)} 
                          \|\D(v^i - v^{i-1})\|_{L^2(\r)},
\eqrn 
where $C$ is an absolute constant and we set 
\[ S = \sup_{i\geq 1} \left(\int_{\r} \vi |\nabla (v^i+f) |^2 dx \right)^{\frac{1}{2}} \]
as in Corollary \ref{S-b}.
Applying Young's inequality, we have 
\bbs 
\int _{\r} [ \ III - V \ ] dx &\geq& - C q^4 (S^2 + S^4) \| v^i - v^{i-1} \|_{L^2(\r)}^2  
                                    \nonumber \\
                             \label{III}
                  &&   - C q^2 S \|\nabla(v^i - v^{i-1})\|_{L^2(\r)}^2 
\nonumber \\
           &&            - \frac{1}{4} \|\D(v^i - v^{i-1})\|_{L^2(\r)}^2
\ees
Combining with (\ref{I}), (\ref{II}) and (\ref{III}),
(\ref{Ediff}) becomes 
\bqrn 
\E(v^{i-1}) - \E(v^i) 
      &\geq& \frac{1}{4}\|\D(v^i - v^{i-1})\|_{L^2(\r)}^2
             + (d\k^2 - C ) \| v^i - v^{i-1} \|_{L^2(\r)}^2 \\
        &&   + (d - C) \|\nabla(v^i - v^{i-1})\|_{L^2(\r)}^2,
\eqrn
where $C$ is an absolute constant depending on $q$ and $\su n_j$.
Taking $d$ large enough, and using the Calderon-Zygmund inequality, we have 
finally
\[ 
\E(v^{i-1}) - \E(v^i) \geq C \|v^i -v^{i-1}\|_{H^2(\r)},
\]
which is a stronger form of  (\ref{E}).
This completes the proof of Lemma 3. \qed
\ \\
\ \\
\begin{Corollary}
Let $v$ be any admissible topological solution of (\ref{veq})-(\ref{Aeq}),
and $v_a ^q$ be the finite energy solution of the Abelian Higgs system.
Then, we have 
\[ \E (v)\leq \E (v_a ^q ). \]
\end{Corollary}
\begin{pf} 
Just substitute $v^i =v, v^{i-1} =v_a ^q $ in the proof of Lemma 3, and 
instead of Lemma 4 we use
\[ \int_{\r} \left[ |1-e^{v + f} + \frac{\k}{q} A_0 | \right]dx  
=\frac{1}{2q^2} \int_{\r} g = \frac{2\pi}{q^2} \su n_j, \]
which follows immediately 
from integration of (\ref{veq})  and Proposition 1.
\end{pf}

\section{Existence of Admissible Solutions and Asymptotic Decay}

Based on the previous estimates for the iteration sequence $\{ v^i \}$,
in this section, we prove
the existence of admissible topological solutions of our Bogomol'nyi equations
(\ref{veq})-(\ref{Aeq}).
We also establish asymptotic exponential decay estimates of these solutions
as $|z| \to \infty $.
As a corollary of these decay estimates we prove that 
the action (\ref{action})  and, hence  the energy functional (\ref{ener}) are finite. 
Firstly we prove
\begin{Thm} 
Given $z_j\in\r$, $n_j\in {\bf Z}_+$ with $j = 1,\cdots , m$, 
there exists a smooth solution $(\phi, A, N)$ to 
(\ref{Bogo1})-(\ref{Bogo4}) such that $\phi = 0$ at each 
$z = z_j$ 
with corresponding winding numbers $n_j$, and satisfying 
\bb
0\leq 1-|\phi|^2 \leq \frac{\k}{q}N =-\frac{\k}{q}A_0
\ee
for all $q, \k >0$
\end{Thm}
\begin{pf} 
By (\ref{E}) the monotone decreasing sequence $\{v^i\}$ satisfies
\[    \E(v^i) \leq \E(v^0) < \infty         \ \ \ \forall i=1,2, \cdots .  \]
This implies by (\ref{coer}) and (\ref{func}) that 
\bb     \label{vkq}
   \| v^i \|_{H^2(\r)} < C \E(v^i) \leq C \E(v^0)  
\ \ \ \forall i=1,2, \cdots .     
\ee
Thus
\[ \sup _{i\geq 0}\| v^i \|_{H^2(\r)} <\infty. \]
On the other hand,
from  (\ref{vit})
\[ A_0^i = \frac{1}{2q\k}\left[ 2q^2 (e^{v^i + f} - 1) + g 
                 + d(v^{i+1}-v^i) - \D v^{i+1}\right]    \] 
which belongs to $L^2 (\r )$ uniformly by  (\ref{vkq}). \\
Thus, $\sup _{i\geq 0}\|A_0^i\|_{L^2(\r)} < \infty$, 
and $\sup _{i\geq 0}\|\D A_0^i\|_{ L^2(\r)} < \infty$ by (\ref{Ait}).
Combining this with the  Calderon-Zygmund inequality
and the standard interpolation inequality, we obtain 
\[\sup _{i\geq 0}\| A_0^i \|_{H^2(\r)} < \infty \]
Thus there exists $v, A_0 \in H^2(\r)$ and a subsequence 
$(v^i , A_0^i )$  such that 
\[ v^i \to v \ \  \mbox{and} \ \  A_0^i \to A_0  \]
both weakly in
$ H^2(\r)$ and  strongly both in  $H^1_{loc} (\r)$ and in  $ L^{\infty}_{loc} (\r)$
by  Rellich's compactness theorem.
The limits $v,A_0 \in H^2(\r) $ satisfies (\ref{veq})-(\ref{Aeq}) 
in the weak sense, and  by repeatedly 
using the standard linear elliptic regularity result we have 
$v,A_0 \in C^{\infty} (\r).$ Moreover, by construction we have
\bb 
v\leq v^q _a \leq -f, 
\ee
and by Proposition 1
\bb
 \frac{q}{\k}(e^{v+f} -1) \leq A_0 \leq 0.
\ee 
We define
\[ N = -A_0, \ \ \phi =\exp \frac{1}{2} (v+f +i \theta) \]
where $\theta =\sum_{j=1} ^m 2n_j arg(z-z_j)$. For 
for $\alpha =  \frac{1}{2}(A_1 -iA_2)$ and $\partial_z =
 \frac{1}{2} (\partial _1 -i\partial_2 )$ we also define
\[\alpha =i \partial_z ln\overline{\phi}. \]
Explicit computation shows
\bb
A_1=\frac{1}{2}(\pa_2 v + \pa_2 b),  
                   \label{A1} 
\ee
and 
\bb
A_2 =\frac{1}{2} (-\partial_1 v - \pa_1 b), 
             \label{A2}
\ee
where  we set
\[    b = -\su n_j \ln ( 1+ |z-z_j|^2).     \]
Converting our reduction procedure from (\ref{veq})-(\ref{Aeq}) to
the Bogomol'nyi equations (\ref{Bogo1})-(\ref{Bogo4}), we
we find that the fields $A_{\mu}, \phi, N$ $(\mu =0, 1, 2)$ satisfy the 
Bogomol'nyi equations (\ref{Bogo1})-(\ref{Bogo4}). 
In particular (32) follows immediately from (34) and (35). 
\end{pf}

We now establish asymptotic exponential decay estimates for admissible
topological solutions of our Bogomol'nyi equations.
\begin{Thm} 
Let $(\phi, A_0 , N)$ be any admissible topological solution of the Bogomol'nyi 
equations (\ref{Bogo1})-(\ref{Bogo4}).
Suppose  $\epsilon >0$ is given, then there exists $r_0 =r_0 (\e )>0$
 and $C=C_{\e}$ such that
\bbs             \label{lon}
0 \leq 1-|\phi|^2,\; |N|,\; |F_{12}| &\leq& C_{\e} e^{-q(1-\g -\epsilon)^{
\frac{1}{2}}|z|}  \\
|D_{\mu}\phi|,\;|\nabla A_0| &\leq& C_{\e} e^{-q(1-\g -\epsilon)^{
\frac{1}{2}}|z|}
                 \label{tra}
\ees
if $|z| > r_0$. Here we set
$\g = \frac{-\rho^2 + \rho\sqrt{\rho^2+8}}{4}$ and 
$\rho = \frac{\k}{q}$.
\end{Thm}
\noindent{\bf Remark:} We note that $\g$ was chosen(see the proof below)
so that $\frac{\rho ^2}{\rho ^2 +2\g} =\g$, thus $0<\g < 1$. \\
\ \\
\noindent{\it Proof of Theorem 3:}
From (\ref{ueq}) and (\ref{rawAeq}),
\bqrn
\D u^2   &=&    2|\nabla u|^2 + 2u \Delta u \\
             &\geq& 4q^2(e^u-1)u-4q\k A_0u, \\
\D A_0^2 &=&    2|\nabla A_0|^2 + 2A_0 \Delta A_0 \\
             &\geq& 2(\k^2 + 2q^2e^u)A_0^2 + 2\k q(1-e^u)A_0
\eqrn
for $|z| > \sup_j \{|z_j|\}$.
Let $E=u^2+2A_0^2$, then we have
\bbs
\D E &\geq& 4q^2\left(\frac{e^u-1}{u}u^2 + 2e^uA_0^2\right) + 4\k^2A_0^2
               +4\k q A_0(1-e^u-u)      \nonumber \\
         &\geq& 4q^2 \xi(u) E + 4\k^2A_0^2 -8\k q A_0 u    \label{decay}
\ees
where we used the inequality $t \leq e^t-1$, and 
set $\xi(t)= \min \{ e^t,\;\frac{e^t-1}{t} \}$. Note that
\bqrn
  \xi(t) &=& e^t \quad \mbox{if} \;\; t< 0 \\
         &=& \frac{e^t-1}{t} \quad \mbox{otherwise}
\eqrn
Also, $\xi >0$ and $\xi(t) \to 1$ as $t \to 0$.
The last term in (\ref{decay}) is estimated by  
\bqrn
8\k q |A_0 u | &\leq& 4(\k^2+ 2\g q^2)A_0^2 
                      + \frac{4\k^2}{\k^2 + 2\g q^2}q^2u^2 \\
               &=& 4(\k^2+ 2\g q^2)A_0^2 + \frac{4\rho^2}{\rho^2 + 2\g}q^2u^2
                =  4\k^2 A_0^2 +  4 \g q^2 E,
\eqrn
since $\frac{\rho^2}{\rho^2 + 2\g}  = \g$.
Thus, (\ref{decay}) becomes
\bb        \label{Edecay}
        \D E \geq 4(\xi(u) -\g)q^2 E        
\ee
Since $u \to 0$ as $|z| \to \infty$, given $\epsilon > 0$, 
we can choose $r_0$ so large that $\xi(u) \geq 1-\epsilon$ on $|z|>r_0$.
Thus, by comparing $E $ with the function $\beta (z)=C_{\e} 
e^{-2q(1-\g -\epsilon)^{\frac{1}{2}}|z|}$ in $|z|\geq r_0$
, using the maximum principle, we deduce 
\[  |u|^2,\;|A_0|^2 \leq C_{\e} e^{-2q(1-\g -\epsilon)^{\frac{1}{2}}|z|}    \]
on $|z| > r_0$, where $C_{\e}$ was fixed to compare $E$ with $\beta$ on 
 $\{|z|=r_0\}$. 
(\ref{lon}) follows from the fact
\[ 0 \leq 1-|\phi|^2 = 1- e^u \leq |u|, \ \ N=-A_0, \] 
and
\[ |F_{12}| \leq q(1-|\phi|^2) + \k |N|, \] 
which follows from (\ref{Bogo3}).
Next, we estimate the asymptotic decay of $|D_{i}\phi|^2$. 
We observe
\bqrn
|D_{\mu} \phi|^2 &=& |(\pa_{\mu} - iqA_\mu)\phi|^2  \nonumber \\
               &=& \frac{1}{4} e^u |(\pa_{\mu} (u +iq\theta) 
                     - iq (\pa_{\mu} ^\perp u + \pa_{\mu} \theta)|^2 \nonumber \\
               &=& \frac{1}{4} e^u |\pa_{\mu} u - iq\pa_{\mu} ^\perp u|^2, 
\eqrn
where we used the notation $(\pa _{\mu} ^\perp )=(\pa _0 , -\pa_2 , \pa _1 )$.
Thus $|D_{\mu}\phi|^2 \leq C|\nabla u|^2$.
Therefore it is sufficient to have decay estimate for
  $|\nabla u|^2$. 
A direct calculation gives
\bqrn
\D |\nabla u|^2 &=&   2|\nabla^2 u|^2 + 2\nabla u \cdot \nabla \D u \\
                &\geq& 4q^2 \nabla u \cdot \nabla (e^u-1-\frac{\k}{q}A_0) \\
                &=&   4q^2 e^u |\nabla u|^2 -4q\k \nabla u\cdot\nabla A_0 \\
\D |\nabla A_0|^2 &=& 2|\nabla^2 A_0|^2 + 2\nabla A_0 \cdot \nabla \D A_0 \\
                  &\geq& 2(\k^2 + 2q^2e^u)|\nabla A_0|^2 
                      - ( 2q\k e^u - 4q^2 e^u A_0) \nabla A_0 \cdot 
                         \nabla u 
\eqrn
for $|z| > \sup_j \{ |z_j| \}$.
We set $J = |\nabla u|^2 + 2 |\nabla A_0|^2$, then  we have  
\bbs 
\D J &\geq& 4q^2 e^u J + 4\k^2 |\nabla A_0|^2 - (8\k q + 4q^2 e^u |A_0|)
                          |\nabla u||\nabla A_0| \nonumber \\
     &\geq&  4q^2 e^u ( 1-\frac{1}{2}|A_0|) J + 4\k^2 |\nabla A_0|^2
 -8\k q|\nabla u||\nabla A_0|, \label{J}
\ees 
where we used $|\nabla u||\nabla A_0| \leq 1/2 J$ in the second inequality.
(\ref{J}) is the same form as (\ref{decay}), observing in case of (\ref{J}) 
we have
\[ e^u ( 1-\frac{1}{2}|A_0|) \rightarrow 1 \ \ \ \mbox{as} \ \ 
|z| \rightarrow \infty. \]
Given $\epsilon > 0$, we apply Young's inequality to the term 
$|\nabla u||\nabla A_0|$ similarly to the previous case, and get
\[ \D J \geq 4q^2 (1-\g -\epsilon) J  \]
when $|z| > r_0$ for  $r_0$ large enough.
The above equation is the same as (\ref{Edecay}). 
Thus $J$ satisfies the estimate (\ref{tra}). \\
This completes the proof of Theorem 5. \qed 
 \ \\ 
Now we complete proof of our existence theorem by proving that the solutions
constructed in Theorem 2 make our action in (2)  finite.
This follows if we prove that any admissible topological solution makes                               
the action finite.
\begin{Corollary}
Let $(A, \phi , N)$ be the solution of the Bogomol'nyi equations (\ref{Bogo1})-
(\ref{Bogo4})
constructed Theorem 2, then  we have
\[  {\cal A} ={\cal A} (A, \phi , N ) < \infty. \]
\end{Corollary}
\begin{pf}
Since $N = -A_0 \in H^1 (\r )$ and $|\phi| \leq 1$,
\[  \int _{\r} (\pa_{\mu} N)^2 dx < \infty,\quad \int_{\r} N^2|\phi|^2 dx < 
\infty.   \]
From (4) and (10)  we have
$F_{12} = q|\phi|^2 -\k N -q  \in L^2 (\r )$.
Clearly
\[   F_{0i} = -\pa_{i} A_0 \in L^2(\r), \quad i = 1,\;2 .  \]
Therefore 
\[  \int_{\r} |F_{\mu \nu}|^2 dx< \infty.                          \]
We now consider the Chern-Simons term.  Firstly we have 
\[ \intr |F_{12} A_0|dx \leq \left( \intr |F_{12}|^2dx \right)^{\frac{1}{2}}
\left( \intr|A_0|^2 dx \right)^{\frac{1}{2}} <\infty .\]
Thus it suffices to prove
\bb 
F_{01} A_2 = -\pa_1 A_0 A_2,\quad F_{02} A_1 = \pa_2 A_0 A_1 \in L^1 (\r). 
\label{f01}
\ee
Since $A_1, A_2 \in L^p (\r)$ for all $p\in (2, \infty]$ from (\ref{A1})-
(\ref{A2}), and 
  $\nabla A_0 \in L^q (\r)$, for $ q \in [1, \infty]$ from (\ref{tra}),
(\ref{f01}) follows immediately by the H\"{o}lder inequality.
Therefore we also have that
\[ \e ^{\mu \nu \l}F_{\mu \nu} A_{\l} \in L^1 (\r ). \]
Finally from  (\ref{tra}) we have 
\[
|D_\mu \phi|^2 \in L^1 (\r). 
\]
This completes the proof of the corollary.
\end{pf}
\section{Abelian Higgs Limit}

In this section we prove that, for $q$ fixed, 
the sequence of admissible topological solutions, $(\v, \A)$
converges to $(v_a ^q ,0)$, as $\k $ goes to zero, where $v_a ^q $ is 
 the finite energy solution of the Abelian Higgs system. 
Firstly we establish:
\begin{Lemma}     \label{uni}
Let $(\v, \A)$ be any admissible topological solution of 
(\ref{veq}) and (\ref{Aeq}). 
Then, for each fixed $q \in (0, \infty)$, we have 
\bb 
\sup_{0<\k < 1} \| \v \|_{H^2(\r)}<\infty, \sup_{0<\k < 1}\|\A\|_{H^2(\r)}. 
 < \infty \label{unik} 
\ee
Thus, by the Sobolev embedding we have
\bb
\sup_{0<\k < 1}\| \v \|_{L^{\infty} (\r )} <\infty  ,\sup_{0<\k < 1}
\|\A\|_{L^{\infty} (\r )}  < \infty \label{vuni} 
\ee
\end{Lemma}
\begin{pf}
Let $\k \in (0, 1)$.
From Corollary 2 and (\ref{coer}) we have
\bqrn
\| \v \|_{H^2(\r)} &\leq &  (1+\k^2)C_1 + C_2 \E (\v )  \\
    &\leq&  (1+\k^2)C_1 + C_2 \E(v^q _a) \leq C (1+\k^2),      
\eqrn
where $C_1, C_2$ and $C$ are  constants independent of $\k$. 
Thus the first inequality of (\ref{unik}) follows.
Now, taking $L^2 (\r )$ inner  product  (\ref{Aeq}) with $\A$, 
we have after integration by part 
\bqrn 
\lefteqn{\int_{\r} |\nabla \A|^2 + (\k^2 + 2q^2e^{\v+f})|\A|^2 \; dx 
           = \int_{\r} \k q(1-e^{\v+f})|\A| \; dx }\hspace{.4in} \\
& \leq & \frac{\k^2}{2} \int_{\r} |\A|^2 dx + \frac{q^2}{2}
\int_{\r}(1-e^{\v+f})^2 \; dx  \\
&\leq & \frac{\k^2}{2} \int_{\r} |\A|^2 dx + \frac{q^2}{2}
\int_{\r} |\v+f|^2 \; dx.  
\eqrn
Thus, by Young's inequality and the first inequality in (\ref{unik}) we obtain
\bb
\int_{\r}|\nabla \A|^2 dx +\intr(\k^2 +4q^2 e^{\v+f})
|\A|^2 dx \leq C, \label{aa1} 
\ee
where $C$ is independent of $\k $.
From (\ref{Aeq}) and (\ref{aa1}) 
 we have
\bqrn
\lefteqn{\int_{\r} |\D \A|^2 \; dx \leq 2\k^2 q^2 \intr (1-e^{\v+f})^2 \; dx 
                      + 2 \intr 
(\k^2 + 2q^2e^{\v+f})^2|\A|^2 \; dx }\hspace{.2in}   \\
                  &\leq&  2\k^2 q^2 \int_{\r} |\v+f|^2 e^{2(\l +f)}\; dx 
                      + 2 (\k^2 +2q^2) \int_{\r} (\k^2 +4q^2 e^{\v+f})
|\A|^2 \; dx \\
 &\leq& 2q^2 \int_{\r} |\v+f|^2 dx + 2(1+q^2) \int_{\r}
 (\k^2 +4q^2 e^{\v+f})|\A|^2  \; dx  \leq C
\eqrn
for a constant $C$ independent of $\k $,
where $\l \in (\v +f, 0)$, and we used the mean value theorem.
Thus, by the Calderon-Zygmund inequality
\bb
\int_{\r} |D^2 \A|^2dx \leq C. \label{aa2} 
\ee
By Sobolev's embedding for a bounded domain, for any ball $B_R=\{|z|<R\}\subset \r$, we have.
\[  \|\A \|_{L^{\infty} (B_R )} \leq C, \] 
where $C$ is independent of $\k $.
We take $R> \max_{1\leq j\leq m} \{ |z_j |+1\}$. Then,
\bbs
\lefteqn{\int_{\r} |\A|^2 dx =  \int_{B_R}  |\A|^2 dx +\int_{\r -B_R}  |\A|^2 dx
}\hspace{.1in}\nonumber \\
 &\leq & \pi R^2  \|\A \|_{L^{\infty} (B_R )}^2 
+ \|e^{\v+f}\|_{L^{\infty} (
\r -B_R)} \int_{\r -B_R}e^{\v+f}  |\A|^2 dx \nonumber \\
&\leq & C_1 (R) + C_2  \|e^f\|_{L^{\infty} (
\r -B_R)}\int_{\r}e^{\v+f}|\A|^2 dx \leq C(R), \label{aa3} 
\ees 
where $C_1 (R), C_2 $ and $C(R)$ are independent of $\k $.
Combining (\ref{aa1})-(\ref{aa3}), we obtain the 
second inequality in (\ref{unik}).
This completes the proof of Lemma 5.
\end{pf}

\begin{Thm} 
Let $\v, \A$ be the admissible topological solutions of (\ref{veq})
and (\ref{Aeq}), and $v_a^q$ a finite energy solution of the Abelian Higgs   
system.
Let $q$ be fixed.  For all $k \in  Z_+$ we have
\[ \v \to v_a^q, \ \ \ \mbox{and}  \quad \A \to 0 \quad \mbox{in}\quad H^k(\r).  \]
as $\k \to 0$.
\end{Thm}
\begin{pf} 
We have by mean value theorem
\[
\D (\v -v_a^q)=2q^2 (e^{\v +f} -e^{v_a^q +f} ) -2\k q \A
\]
\bb
=2q^2 e^{\l +f} (\v -v_a^q) -2\k q \A \label{diff} 
\ee
where $\l \in (\v, v_a^q )$.
Multiplying (\ref{diff}) by $\v-v_a^q$, we have after
integration by parts
\[          
\int_{\r} |\nabla (\v-v_a^q)|^2 + 2q^2 e^{\l+f} (\v-v_a^q)^2 \; dx 
         = 2\k q \int_{\r} \A (\v -v_a^q) \; dx
\]
\[
\leq 2\k q \|\A\|_{L^2(\r)} \|\v -v_a^q\|_{L^2(\r)}
 \leq \k C
\]
where   
 $C$ is independent of $\k$ by Lemma \ref{uni}.  Since 
\[
\|\l \|_{L^{\infty} (\r )} \leq
\|\v \|_{L^{\infty} (\r )}+\|v_a^q\|_{L^{\infty} (\r )}  \leq C 
\]
independently of $\k< 1$,
we have from the above estimate
\[
\int_{\r} |\nabla(\v- v_a^q)|^2 + e^f |\v -v_a^q|^2 \; dx \to 0
\]
as $\k \to 0$.
Let $\Omega_{\delta}=\cup_{j=1} ^m \{ |z-z_j|< \delta\}$.
Now,
\[\int_{\Omega_{\delta}} |\v -v_a^q |^2 dx \leq
\pi m\delta^2 \|\v -v_a^q\|_{L^{\infty} (\r )} ^2 \leq C\delta^2. \] 
where   
 $C$ is independent of $\k$ by Lemma \ref{uni}.
Thus, for any given $\e >0$, we can choose $\delta$ independently of
$\k$ so that
\[ \int_{\Omega_{\delta}} |\v -v_a^q |^2 dx  \leq \frac{\e}{2}.\]
For such $\delta$ we have
\bqrn
\int_{\r}  |\v -v_a^q |^2 dx &=& \int_{\Omega_{\delta}} |\v -v_a^q |^2 dx
+\int_{\r-\Omega_{\delta}} |\v -v_a^q |^2 dx \\
&\leq& \frac{\e}{2} +\sup_{\r-\Omega_{\delta}}\{e^{|f|}\} 
\int_{\r} e^f|\v -v_a^q |^2 dx \\
&\leq& \frac{\e}{2} +\frac{\e}{2} =\e 
\eqrn
for sufficiently small $\k$, i.e.
\[\int_{\r}  |\v -v_a^q |^2 dx \to 0 \]
as $\k \to 0$.
Combining the above results, we obtain
\[
\v \to v_a^q \  \ \mbox{in} \ \ H^1 (\r ) \ \ \mbox{as} \ \ \ \k\to 0.
\] 
Now we prove the convergence for $\A$. 
Multiplying (\ref{Aeq}) by $\A$ and integrating, we estimate
\[
\int_{\r} |\nabla \A|^2 + (\k^2 + 2q^2 e^{\v+f}) |\A|^2 \; dx
                  \leq  \k q\int_{\r} (1 - e^{\v+f}) |\A| \; dx
\]
\[                  \leq \k q\|\A\|_{L^2(\r)} \|1-e^{\v+f}\|_{L^2(\r)}
                  \leq C \k \|\v+f\|_{L^2(\r)} \leq C \k.
\]
where we used Lemma \ref{uni} in the first and third step and 
use the fact $1-e^t \leq t$ for $t \leq 0$ in the second step. 
Using the fact $|\v| < C$ uniformly in $\k <1$, we obtain from this
\[ \int_{\r} |\nabla \A|^2 + e^f |\A|^2 \; dx \to 0 \]
as $\k \to 0$. Since $|\A| <C$ uniformly in $\k < 1$, by Lemma \ref{uni}
we can deduce 
$\A \to 0$ in $H^1(\r)$ similarly to  the case of $\v$.
From these results together with 
uniform bounds $\|\v\|, \|\A\| \leq C$, applying 
the standard elliptic regularity  to  
(\ref{diff}) and (\ref{Aeq}) repeatedly, we obtain
\[ (\v,\; \A )\to (v_a^q,\; 0) \quad \mbox{in} \quad [H^k (\r)]^2,\quad
  \forall k \geq 1                    \]
\end{pf}

\section{Chern-Simons Limit} 

In this section we study the behaviors of $\v, \A$ as 
$\k , q \to \infty$ with $l=q^2/\k$ kept fixed for the admissible
topological solutions.
Although we could not obtain the strong convergence to a solution
of the Chern-Simons equation, instead, we will prove that the sequence
$\{ \v\}$ is "weakly approximating"  the  Chern-Simons equation:
\[ \D v =4l^2e^{v+f} (e^{v +f} -1) +g. \] 
We denote $l=q^2/\k$ the fixed number,  and $\tA=q \A$ throughout this section.  
\begin{Thm}\label{chern}
Let $ \{(\v , \A )\}$ be a sequence of  admissible topological
solutions of (\ref{veq})-(\ref{Aeq}). 
For any $\psi \in C_0 ^{\infty} (\r )$ we have
\bb
\Lint \left[ \D \v  - 
4l^2 e^{\v+f} (e^{\v +f} -1) -g\right] \psi dx =0.
\ee 
\end{Thm}
For proof of this theorem we firstly establish the following lemma
which is interesting in itself.
\begin{Lemma}
Let $ \{(\v , \A )\}$ be given as in Theorem 5.
For any fixed $p\in [1, \infty)$ we have
\[
\Li \| \tA -l (e^{\v +f} -1)\|_{L^p (\r)} =0.
\]
\end{Lemma}
\begin{pf}
Firstly we have from $ \v +f \leq 0$
\[ \|e^{\v +f} -1\|_{L^{\infty} (\r)} \leq 1.\] 
Also, from $0\geq \tA  \geq l(e^{\v +f} -1)$,
\[ \|\tA\|_{L^{\infty} (\r)}\leq l\|e^{\v +f} -1\|_{L^{\infty} (\r)}
\leq l. 
\]
From (\ref{veq}) we have
\bqrn 
\int_{\r} |\tA -l(e^{\v +f} -1)|dx &=&
\int_{\r} \left[\tA -l(e^{\v +f} -1)\right]dx \\
&=& \frac{1}{q} \int_{\r} g dx.
\eqrn
Thus
\[ \Li \| \tA -l (e^{\v +f} -1)\|_{L^1 (\r)} =0.\]
By a standard interpolation inequality 
\bqrn 
\lefteqn{\| \tA -l (e^{\v +f} -1)\|_{L^p (\r)}}\hspace{ .2in} \\
&\leq&\| \tA -l (e^{\v +f} -1)\|_{L^1 (\r)}^{\frac{1}{p}}
\| \tA -l (e^{\v +f} -1)\|_{L^{\infty} (\r)}^{1-\frac{1}{p}} \\
&\leq& (2l)^{1-\frac{1}{p}}\| \tA -l (e^{\v +f} -1)
\|_{L^1 (\r)}^{\frac{1}{p}}
 \to 0
\eqrn
as $\k, q \to \infty$ with $q^2/\k =l $ fixed.
\end{pf}
\ \\
\noindent{\it Proof of Theorem 5:} 
From (13) added by (14)$\times q/\k$ we obtain
\[ 
\D (\v +\frac{2l}{q} \tA )=g+4l e^{\v +f} \tA. 
\]
Multiplying $\psi \in C_0 ^{\infty} (\r )$, and integrating by parts,
we obtain
\[
\int_{\r}  \D \v \psi =4l^2 \int_{\r}\left[ e^{\v+f} 
(e^{\v +f} -1) +g\right] \psi dx
\]
\[
+ \frac{2l}{q}\int_{\r} \tA\D\psi dx +4l \int_{\r} e^{\v+f} [
\tA -l(e^{\v +f} -1)]\psi dx.
\]
Now we have
\[ \Li\left|\frac{2l}{q}\int_{\r} \tA\D\psi dx\right|
\leq \Li\frac{2l^2}{q} \int_{\r}|\D \psi|dx =0.
\]
and by  Lemma 6 
\[
\Lint e^{\v +f}[\tA -l(e^{\v +f} -1)]dx
\]
\[
\leq
 \|\psi\|_{L^{\infty} (\r )}\Lint |\tA -l(e^{\v +f} -1)|dx =0.
\]
Thus, Theorem 5 follows. \qed \\
\ \\
\noindent{\bf Remark:} If we could have uniform $L^1 (\r)$ estimate
of $\nabla \v$, then we could prove existence of
subsequence $\{\v\}$ and its $L^q _{loc} (\r)$ $(1\leq q< 2)$-limit 
$v$ such that $v$ is a smooth solution of 
the Chern-Simons equation. 
 \ \\
\[ \mbox{\bf Acknowledgements } \]
The authors would like to deeply 
thank to Professor Choonkyu Lee for introducing
the problems issued in this paper for them, and many helpful discussions. 
This research is supported partially by KOSEF(K94073, K95070),
BSRI(N94121), GARC-KOSEF and SNU(95-03-1038).


\begin{thebibliography}{99}
\bibitem{D} G. Dunne, {\it Self-Dual Chern-Simons Theories,}
             Springer Lecture Note in Physics, M36,(1995).
\bibitem{H} J. Hong, Y. Kim and P. Y. Pac, Phys. Rev. Lett. {\bf 64},
                  pp. 2230, (1990). 
\bibitem{J} R. Jackiw and E. J. Weinberg, Phys. Rev. Lett.  {\bf 64}, 2234,
               (1990)
\bibitem{BLee} B. H. Lee, C. Lee and H. Min. 
               Phys. Rev. D, {\bf 45}, pp. 4588 (1990).
\bibitem{Lee} C. Lee, K. Lee and H. Min, {\it Self-Dual Maxwell 
                 Chern-Simons solitons}, Phys. Lett. B, {\bf 252},
                 pp. 79-83(1990).
\bibitem{Lee1} C. Lee, K. Lee and E. J. Weinberg,  Phys. Lett. B
                {\bf 243}, pp. 105- (1990).
\bibitem{SY} J. Spruck and Y. Yang, {\it Existence Theorems for Periodic
            Nonrelativistic Maxwell-Chern-Simons Solitons}, to appear in
                J. Diff. Eqns. (1996).
\bibitem{Spruck} J. Spruck and Y. Yang, {\it Topological 
                 Solutions in the Self-Dual Chern-Simons Theory:
                 Existence and Approximation}, Ann. Inst. Henri
                 Poincar\'{e}, {\bf 12}, pp. 75-97(1995)
\bibitem{Spruck1} J. Spruck and Y. Yang, {\it The Existence of Non-Topological
        Solitons in the Self-Dual Chern-Simons Theory},
        Comm. Math. Phys. {\bf 149}, pp. 361-376(1992).
\bibitem{Taubes} A. Jaffe and C. Taubes, Vortices and Monopoles,
             Birkh\"{a}user, Boston, 1980.
\bibitem{Wang} R. Wang, {\it The Existence of Chern-Simons Vortices},
                  Comm. Math. Phys., {\bf 137}, pp. 587-597(1991)
\bibitem{WangY} S. Wang and Y. Yang, {\it Abrikosov's Vortices in the 
Critical Coupling}, SIAM J. Math. Anal. Vol. {\bf 23}, 1992, pp. 1125-1140.
\end{thebibliography}
\end{document}